\documentclass[times, twoside]{zHenriquesLab-StyleBioRxiv}
\usepackage{booktabs}

\leadauthor{Jie Xiao}

\begin{document}

\title{A Two-Stage Strategy for Mitosis Detection Using Improved YOLO11x Proposals and ConvNeXt Classification}
\shorttitle{Two-Stage Mitosis Detection for MIDOG 2025}

\author{Jie Xiao}
\author{Mengye Lyu}
\author{Shaojun Liu} 

\affil{College of Health Science and Environmental Engineering, Shenzhen Technology University, Shenzhen, 518118, China}

\maketitle

\begin{abstract}
MIDOG 2025 Track 1 requires mitosis detection in whole-slide images (WSIs) containing non-tumor, inflamed, and necrotic regions. Due to the complicated and heterogeneous context, as well as possible artifacts, there are often false positives and false negatives, thus degrading the detection F1-score. To address this problem, we propose a two-stage framework. Firstly, an improved YOLO11x, integrated with EMA attention and LSConv, is employed to generate mitosis candidates. We use a low confidence threshold to generate as many proposals as possible, ensuring the detection recall. Then, a ConvNeXt-Tiny classifier is employed to filter out the false positives, ensuring the detection precision. Consequently, the proposed two-stage framework can generate a high detection F1-score. Evaluated on a fused dataset comprising MIDOG++, MITOS\_WSI\_CCMCT, and MITOS\_WSI\_CMC, our framework achieves an F1-score of 0.882, which is 0.035 higher than the single-stage YOLO11x baseline. This performance gain is produced by a significant precision improvement, from 0.762 to 0.839, and a comparable recall. On the MIDOG 2025 Track 1 preliminary test set, the algorithm scores an F1 score of 0.7587. The code is available at \href{https://github.com/xxiao0304/MIDOG-2025-Track-1-of-SZTU}{https://github.com/xxiao0304/MIDOG-2025-Track-1-of-SZTU}.
\end{abstract}


\begin{corrauthor}
liusj14@tsinghua.org.cn 
\end{corrauthor}

\section*{Introduction}
Mitotic figure counting is critical for tumor grading in clinical pathology. However, manual annotation of mitosis is time-consuming due to the extra-large size of whole slide images (WSIs). Besides, the morphology of mitotic figures strongly overlaps with similar-looking impostors, leading to annotation variability among pathologists~\cite{ammeling_mitosis_2025}. MIDOG 2025 Track 1 focuses on mitosis detection on \textit{full WSIs} instead of localized regions of interest, introducing three key technical hurdles: 
\begin{itemize}
    \item Small mitotic figures, typically 10–30 pixels, are easily missed in low-resolution WSI patches; 
    \item Necrotic debris and inflamed cells mimic the morphological features of mitoses, leading to high false positives; 
    \item The dataset includes 12 tumor types of human and veterinary specimens. Such domain shifts might reduce the generalization of feature-based models.  
\end{itemize}

Considering the large size of WSIs, the model should be efficient. Single-stage detectors such as YOLO11x~\cite{khanam2024yolov11} prioritize inference efficiency; therefore, they are suitable for this task. However, they usually struggle with false positive suppression in hard regions, including necrotic zones. Indeed, we can decrease the false positives by using a lower confidence threshold. However, this would lead to more false negatives, consequently degrading the detection F1-score.
\begin{figure}[!t]  
    \centering
    \includegraphics[width=0.5\textwidth]{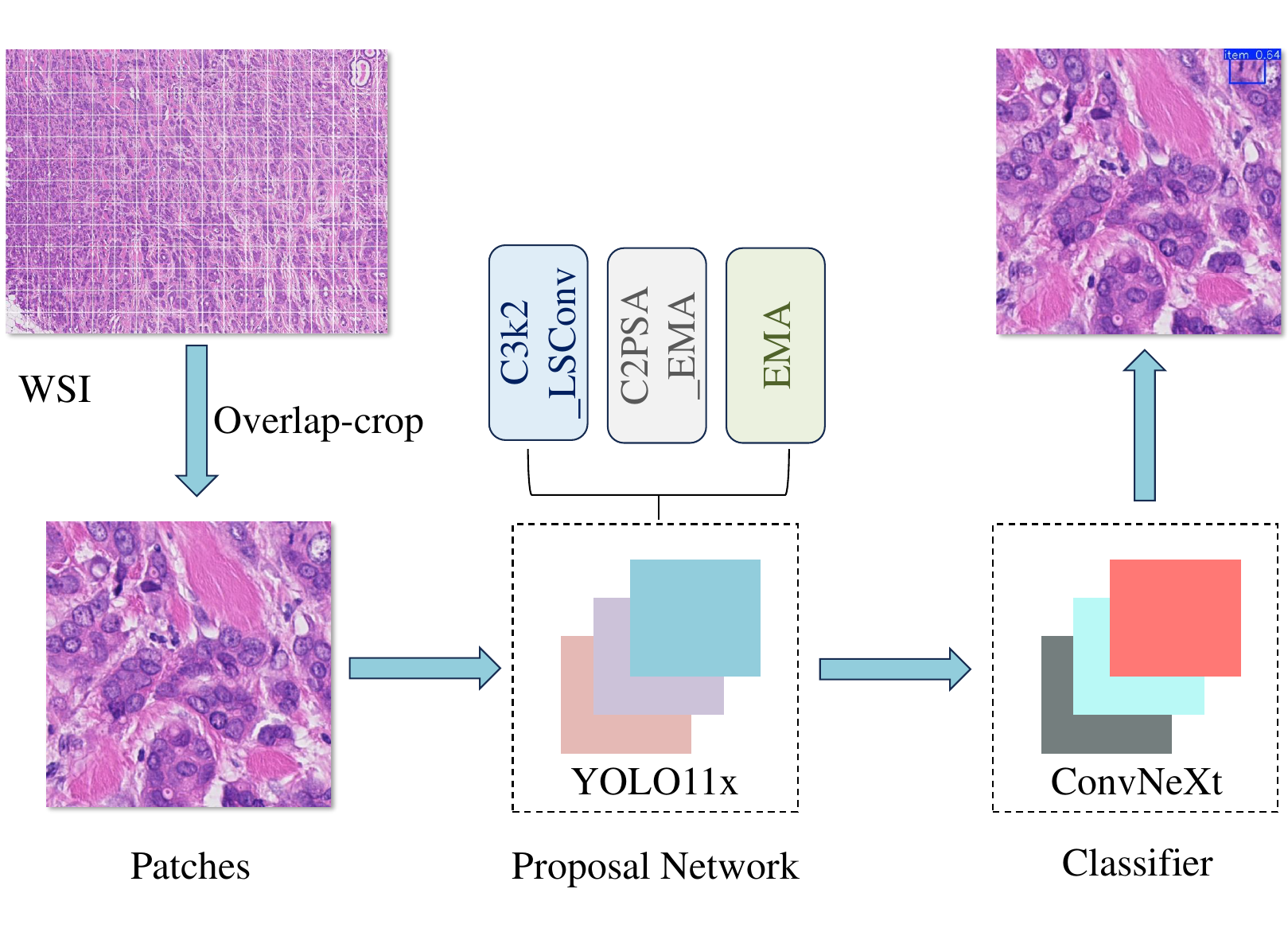}  
    \caption{Two-Stage Mitosis Detection Framework Workflow. 
             The pipeline includes patch cropping, candidate generation with an improved YOLO11x~\cite{khanam2024yolov11} as proposal network, and false positive filtering with a ConvNeXt~\cite{liu2022convnet2020s} as classification network.}
    \label{fig:two_stage_framework} 
\end{figure}

To tackle this problem, we design a two-stage pipeline comprising an improved YOLO11x and a classification network. Specifically, a \textit{recall-oriented improved YOLO11x proposal network}, with a low confidence threshold, is employed to capture all potential mitoses first; and the candidates are filtered with a \textit{precision-oriented ConvNeXt classifier} \cite{liu2022convnet2020s} to reduce false positives. This design can keep the high inference efficiency of YOLO11x while improving the detection F1-score. This synergy between YOLO11x’s efficiency and ConvNeXt’s robustness directly addresses the unique challenges of MIDOG 2025 Track 1.

\section*{Material and Methods}
\subsection*{1. Datasets and Preprocessing}
Three publicly available data sets are used for this study, with data divided into training, validation, and test sets in a ratio of \textbf{7:1:2}. All data usage complies with the MIDOG 2025 Track 1 guidelines~\cite{ammeling_mitosis_2025}. Detailed information about each dataset is as follows:

\begin{itemize}
    \item \textbf{MIDOG++} contains 553 patches sized 7,200$\times$5,400, from 503 tumor cases spanning 7 tumor types, with 11,937 manually annotated mitotic figures~\cite{ammeling_mitosis_2025}.
    
    \item \textbf{MITOS\_WSI\_CMC} includes 21 WSIs of canine breast cancer with 13,907 mitotic annotations~\cite{aubreville_completely_2020}.
    
    \item \textbf{MITOS\_WSI\_CCMCT} comprises 32 WSIs of canine mast cell tumors with 44,880 annotations, featuring abundant necrotic and inflamed regions~\cite{aubreville_comprehensive_2023}.
\end{itemize}

Since either the patches in MIDOG++ or the WSIs in the rest two datasets are too large for the model to process, we crop them into 512$\times$512 local patches with 20\% overlap to preserve the integrity of mitotic figures, avoiding partial mitoses at the patch edges.

To ensure reliable and unbiased evaluation, data augmentation is applied exclusively to the training set; no augmentation is performed on the validation or test sets. The augmentations include:
\begin{itemize}
    \item Geometric transformations: random rotation in the range of $[0\degree,180\degree]$; random horizontal and vertical flip with a probability of 0.5.  
    \item Mixing strategies: random mixup with another local patch, with a probability of 0.3 and a random weight in the range of $[0,1]$; mosaic with another 3 random local patches, which is only utilized in the first 20 epochs to stabilize late-stage training.
    \item Advanced augmentation: RandAugment~\cite{cubuk2019randaugmentpracticalautomateddata} with default settings; random erasing with a probability of 0.4 and an erase ratio in the range of $[0.02, 0.1]$.
\end{itemize}

\begin{figure}[!t]  
    \centering

    \includegraphics[width=0.5\textwidth]{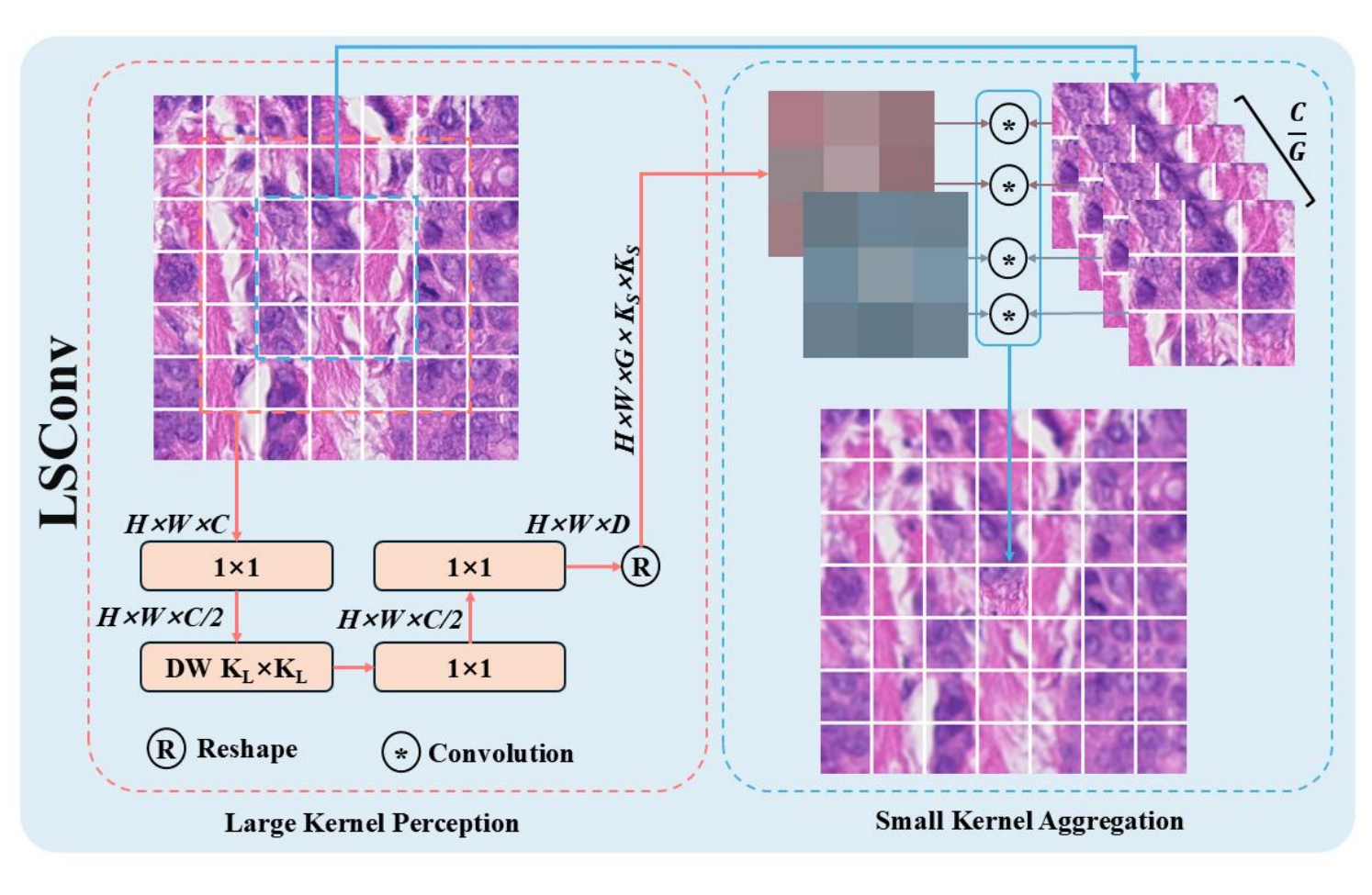}  
    \caption{Network Architecture of LSConv~\cite{wang2025lsnetlargefocussmall}. 
         LSConv combines large-kernel perception and small-kernel aggregation. It simulates the dynamic multi-scale visual capability of the human visual system. Large-kernel perception utilizes large-kernel depth-wise convolution to capture extensive contextual information, while small-kernel aggregation performs fine-grained aggregation of features within a small range through dynamic convolution.}
    \label{fig:lsconv_architecture}  
\end{figure}

\subsection*{2. Two-Stage Framework Design}
The framework is illustrated in Fig.~\ref{fig:two_stage_framework}. Firstly, the WSIs are cropped into local patches, and an improved YOLO11x proposal network is performed under a low confidence threshold to generate mitosis candidates, reducing false negatives. Then, the candidates are filtered by a ConvNeXt classifier network to reduce false positives. Therefore, such a two-stage design can ensure both recall and precision, leading to a high F1-score. The details are described as follows.
\subsubsection*{Stage 1: Proposal Network (YOLO11x + LSConv + EMA)}
The base model architecture follows YOLO11x design principles~\cite{khanam2024yolov11}. To enhance small-target recall while reducing model complexity, the original YOLO11x is improved with three key modules:
\begin{itemize}
    \item \textbf{C3k2\_LSConv Blocks.} The C3k2 blocks in the P3, P4, and P5 detection heads are replaced with C3k2\_LSConv blocks. Specifically, the convolution layers in the C3K2 block are replaced with the LSConv layer~\cite{wang2025lsnetlargefocussmall}, whose core architecture is illustrated in Fig.~\ref{fig:lsconv_architecture}. Each LSConv integrates a 7$\times$7 depth-wise convolution to enlarge the receptive field for small targets and a 3$\times$3 dynamic grouped convolution to enhance feature discrimination. This design can improve efficiency while maintaining performance.
    \item \textbf{C2PSA\_EMA Module.} It is deployed at the fusion interface between the backbone and neck. The input is divided into several groups and then sequentially fed into a 1$\times1$ convolution, and $n$ PSABloc\_EMA units, where each unit incorporates a feed-forward network and EMA attention~\cite{Ouyang_2023}. Finally, the groups are combined together to restore the original channel dimension.
    \item \textbf{EMA Attention.} To further refine feature representation at the detection stage, this cross-spatial attention mechanism~\cite{Ouyang_2023} is additionally attached to each detection head, \textit{i.e.}, P3, P4, and P5. It groups channels by a factor of 32 to model dependencies, and through its channel-grouping strategy and cross-spatial interaction, effectively suppresses background noise in necrotic and inflamed regions.
\end{itemize}

  As mentioned above, the proposal network needs to generate as many candidates as possible, aiming to avoid false negatives. Therefore, a low confidence threshold of 0.2 is employed to maximize the recall. Additionally, in the non-maximum suppression procedure, the intersection over union (IoU) threshold is set to 0.3 to reduce redundant candidates.

\subsubsection*{Stage 2: ConvNeXt Classifier}
Since the proposal network is performed under a low confidence threshold, there are many false positives. Therefore, the basic ConvNeXt-Tiny network~\cite{liu2022convnet2020s} is employed to further classify the mitosis candidates to reject false positives.

As the candidates are usually of different sizes, they are resized to 64×64 pixels and normalized using ImageNet statistics. During training, the preprocessed proposals are augmented with the following strategies.
\begin{itemize}
    \item Geometric/color transformations: random horizontal flip with a probability of 0.5; random rotation where the angle is in the range of $[-15\degree, +15\degree]$; ColorJitter where the jitter parameters for brightness, contrast, and saturation are 0.2, 0.2, and 0.1, respectively.;  
    \item  Mixing strategies: random mixup with another candidate, with a probability of 0.2 and a random weight in the range of $[0,1]$.
    \item  Advanced augmentation: RandAugment~\cite{cubuk2019randaugmentpracticalautomateddata} with 3 types of operation whose magnitude is 5; random half erasing with a probability of 0.5 and an erase ratio in the range of $[0.02, 0.15]$.  
\end{itemize}

To train the classifier, we use a hybrid loss with focal loss to address the class imbalance issue, and contrastive loss to enhance feature separation between mitosis hard negatives. The loss function is described as (\ref{equ:total_loss}).

\begin{equation}
    \label{equ:total_loss}
    \mathcal{L}_{\text{total}} = \mathcal{L}_{\text{focal}} + \lambda \cdot \mathcal{L}_{\text{contrastive}}.
\end{equation}
Here, $\lambda$ is a balancing weight between focal loss $\mathcal{L}_{\text{focal}}$, which is described in (\ref{equ:focal_loss}), and contrastive loss $\mathcal{L}_{\text{contrastive}}$, which is described in (\ref{equ:contrastive_loss}).

\begin{equation}
    \label{equ:focal_loss}
    \mathcal{L}_{\text{focal}} = -\frac{1}{N} \sum_{i=1}^{N} \alpha_{c_i} \cdot (1 - p_{c_i})^\gamma \cdot \log(p_{c_i}),
\end{equation}

\begin{equation}
    \label{equ:contrastive_loss}
    \mathcal{L}_{\text{contrastive}} = -\frac{1}{N} \sum_{i=1}^{N} \log\left( \frac{\exp\left(\text{sim}(f_i, f_i^{+})/T \right)}{\sum_{\substack{j=1}}^{N} \exp\left( \text{sim}(f_i, f_j)/T \right)} \right),
\end{equation}
where $N$ is the total number of samples in a batch;
$\gamma$ is the focusing parameter to down-weight easily classified samples;
$c_i$ is the class for $i$-th sample; $\alpha_{c_i}$ is the weight for that class; $p_{c_i}$ is the predicted probability of the $i$-th sample belonging to its true class $c_i$;
$\text{sim}(\cdot, \cdot)$ is the cosine similarity; $f_i$ and $f_j$ are the feature vectors for the $i$-th and $j$-th samples;
$f_i^+$ is the feature vector of a sample whose class is the same as sample $i$.
$T$ is the temperature parameter adjusting the steepness of the similarity distribution.

\section*{Training and Inference}
\subsection*{1. Training Details}
The training is conducted on a server equipped with 8 NVIDIA A100 80G GPUs. The training parameters are set as follows.
\begin{itemize}
    \item \textbf{Proposal Network.} The loss function is defaultly set according to~\cite{khanam2024yolov11}. The network is optimized using the AdamW optimizer with cosine annealing, where the initial learning rate is $10^{-3}$ and the final learning rate is $10^{-5}$, and a weight decay of $5\times10^{-4}$. The model is trained for 300 epochs with a batch size of 960.
    \item \textbf{ConvNeXt-Tiny Classifier.} The parameters in the loss functions are empirically set as: $\alpha_{m}=1, \alpha_{b}=1.5, \gamma=2.0, T=0.2, \lambda=1.0$. The network is optimized using the AdamW optimizer with cosine annealing, where the initial learning rate is $3\times10^{-4}$ and the final learning rate is $10^{-6}$, and a weight decay of $10^{-5}$. The model is trained for 400 epochs with an early stopping strategy monitoring on the validation F1-score, with a patience of 60. The batch size is 960.
\end{itemize}

\subsection*{2. WSI Inference Pipeline}
\begin{enumerate}
    \item WSIs are split into 512$\times$512 local patches with 20\% overlap to ensure continuous coverage;
    \item The proposal network generates candidate mitotic regions with confidence scores $\ge 0.2$;  
    \item Each candidate region is resized to 64$\times$64 pixels, then fed into the ConvNeXt-Tiny classifier; candidates with classification scores $< 0.5$ are rejected;
    \item The remaining candidates are merged across overlapping patches when IoU $\ge 0.5$, to generate the final detection results.  
\end{enumerate}

\subsection*{3. Evaluation Metrics}  
Aligned with the evaluation guidelines of MIDOG 2025 Track 1 \cite{ammeling_mitosis_2025}, F1-score ($F1$), recall ($R$), and precision ($P$) are employed. Their definitions are:
\begin{align}
    \label{equ:F1}F1 = \frac{2*P*R}{P+R},\\
    \label{equ:R} R = \frac{TP}{TP+FN},\\
    \label{equ:P} P = \frac{TP}{TP+FP},
\end{align}
where $TP$, $FP$, and $FN$ are the number of true positives, false positives, and false negatives, respectively.

\section*{Results}
\subsection*{Quantitative Results}
We compare the proposed two-stage framework with two model variants: the basic YOLO11x and the improved YOLO11x. Both are single-stage variants. The quantitative results are summarized in Table~\ref{tab:performance}.

\begin{table}[!htbp]
\centering
\caption{Quantitative results on the test set. The best results are in \textbf{bold}, and the second-best results are \underline{underlined}.}
\label{tab:performance}
\setlength{\tabcolsep}{2pt}  
\resizebox{0.45\textwidth}{!}{
\begin{tabular}{@{}c *6{>{\centering\arraybackslash}p{0.9cm}}@{}}
\toprule
\textbf{Model Variant}      & $\mathbf{TP\uparrow}$    & $\mathbf{FP\downarrow}$   & $\mathbf{FN\downarrow}$   & $\mathbf{P\uparrow}$ & $\mathbf{R\uparrow}$ & $\mathbf{F1\uparrow}$ \\
\midrule
Basic YOLO11x     & \textbf{17879} & 7165 & \textbf{439}  & 0.716     &\textbf {0.976}  & 0.827    \\
Improved YOLO11x       & \underline{17441} & \underline{5433} &\underline {877}  & \underline{0.762}     & \underline{0.952}  & \underline{0.847}    \\
Two-Stage (Ours)   & 17030 & \textbf{3272} & 1288 &\textbf{0.839}      & 0.929  & \textbf{0.882}    \\
\bottomrule
\end{tabular}
}
\vspace{-10pt}
\end{table}

\subsubsection*{1. Basic YOLO11x}  
The basic YOLO11x achieves the highest recall of 0.976, with the fewest false negatives, demonstrating its strong capability in ``broad mitotic capture'', a critical requirement for full-WSI detection. However, it suffers from significantly more false positives, resulting in a precision of only 0.716. This is mainly caused by the misclassification of necrotic debris as mitoses, which limits its clinical utility. Though it has the highest recall, the F1-score is degraded significantly by the low precision.

\subsubsection*{2. Improved YOLO11x}  
With the integration of EMA attention and LSConv, the improved YOLO11x model avoids false positives to some extent. 

FP decreases to 5433, a 24\% reduction with respect to the basic YOLO11x. Consequently, the precision improves to 0.762. 
At the same time, the recall decreases to 0.952. Despite this sacrifice, the F1-score still improves to 0.847, validating the rationality of the "fewer false positives over minor missed detection" design principle. 

\subsubsection*{3. Two-Stage Framework (Ours)}  
Incorporating the ConvNeXt classifier with the improved YOLO11x further enhances precision while maintaining stable recall.  
FP is reduced to 3272, a 40\% reduction with respect to the improved YOLO. Consequently, the precision jumps to 0.839. 
At the same time, the recall decreases to 0.929, which is still clinically viable for mitotic counting.  
Despite this sacrifice, the F1-score reaches 0.882, which is 0.055 higher than the basic YOLO11x, and 0.035 higher than the improved YOLO11x. On the MIDOG 2025 Track 1 preliminary test set,our method achieved an F1 score of 0.7587.
\subsubsection*{Qualitative Results}
\begin {figure} 
\centering
\includegraphics [width=0.5\textwidth]{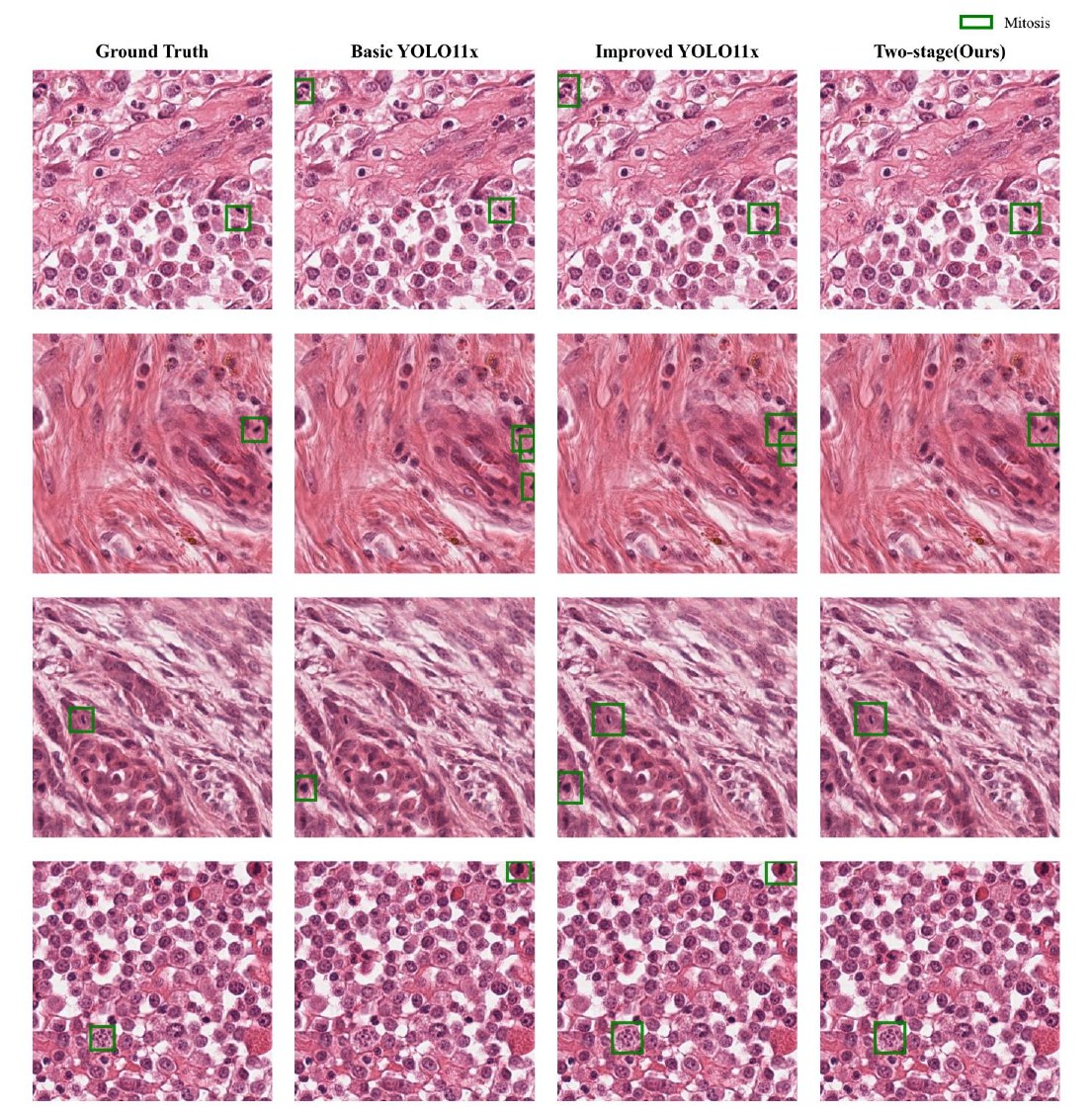} 
\caption {Visual Results of Mitosis Detection.
The figure shows a comparison of detection results among Ground Truth, Basic YOLO11x, Improved YOLO11x, and the proposed Two-stage framework. Green boxes indicate correctly detected mitosis regions. There are many false positives for basic YOLO11x; Improved YOLO11x can partially reduce the false positives; The proposed two-stage method can further reject hard false positives.}
\label {fig:qualitative_results} 
\end {figure}

To visually evaluate the performance of the proposed two-stage framework, we compare it with two single-stage model variants, \textit{i.e.}, the basic YOLO11x and the improved YOLO11x. The results are shown in Fig.~\ref{fig:qualitative_results}. It can be observed that the basic YOLO11x has more false positives compared to the improved YOLO11x, while the two-stage framework reduces false positives even more than the improved YOLO11x. This coincides with the quantitative results.

\section*{Discussion and Conclusion}
\paragraph{Contribution.}
The proposed two-stage ``Improved YOLO11x + ConvNeXt'' framework delivers two key advantages: 
\begin{itemize}
    \item \textbf{Targeted Design.} The ``broad recall + precision filtering'' workflow directly addresses the inefficiency of full-WSI detection, mitigating false positives in necrotic regions.  
    \item \textbf{Superior Performance.} It achieves a higher F1-score, with 54\% reduction in false positives with respect to the basic YOLO11x. 
\end{itemize}

\paragraph{Future Work.}  
Future research will focus on two directions:  
1. Developing a dynamic threshold adjustment mechanism that adapts to WSI-specific features, such as the proportion of necrotic regions, staining intensity, and tumor type;  
2. Integrating domain adaptation modules, \textit{e.g.}, domain-adversarial training, to eliminate feature distribution gaps across datasets, further improving the framework’s generalization to diverse clinical scenarios.  

\section*{Bibliography}
\bibliography{literature} 
\typeout{get arXiv to do 4 passes: Label(s) may have changed. Rerun}
\end{document}